\documentclass[a4paper]{article}
\usepackage{amsmath,amssymb,theorem,fullpage}
\usepackage[nosort]{cite}

\usepackage{graphicx}

\def\eq#1{\begin{equation}#1\end{equation}}
\def\Aligned#1{\begin{aligned}#1\end{aligned}}
\def\p{\partial}

\def\beqn{\begin{eqnarray}}
\def\eeqn{\end{eqnarray}}
\def\nn{\nonumber}
\def\beq{\begin{equation}}
\def\eeq{\end{equation}}
\def \ra {\rangle}
\def \la {\langle}
\def \lb {\left(}
\def \rb {\right)}
\newtheorem{prop}{Proposition}
\newtheorem{thm}{Theorem}
\DeclareMathOperator{\ima}{im}

\allowdisplaybreaks

\numberwithin{equation}{section}

\makeatletter

\def\section{\@startsection{section}{1}{\z@}%
                                   {3.5ex \@plus -1ex \@minus -.2ex}%
                                   {2.3ex \@plus.2ex}%
                                   {\normalfont\normalsize\bfseries}}
\def\subsection{\@startsection{subsection}{2}{\z@}%
                                     {3.25ex\@plus -1ex \@minus -.2ex}%
                                     {1.5ex \@plus .2ex}%
                                     {\normalfont\normalsize\bfseries\itshape}}
\def\subsubsection{\@startsection{subsubsection}{2}{\z@}%
                                     {3.25ex\@plus -1ex \@minus -.2ex}%
                                     {1.5ex \@plus .2ex}%
                                     {\normalfont\normalsize\bfseries\itshape}}
\def\@seccntformat#1{\csname the#1\endcsname.~~}
\long\def\@makecaption#1#2{%
  \vskip\abovecaptionskip
  \sbox\@tempboxa{\small#1. #2}%
  \ifdim \wd\@tempboxa >0.9\hsize
  {\leftskip=0.05\hsize\rightskip=0.05\hsize\relax\small
    #1. #2\par}
  \else
    \global \@minipagefalse
   \hb@xt@\hsize{\hfil\box\@tempboxa\hfil}%
  \fi
  \vskip\belowcaptionskip}
\def\Appendix{\appendix
  \def\@seccntformat##1{Appendix~\csname the##1\endcsname.~~}}

\let\over\@@over
\let\atop\@@atop
\let\above\@@above
\let\overwithdelims\@@overwithdelims
\let\atopwithdelims\@@atopwithdelims
\let\abovewithdelims\@@abovewithdelims

\makeatother

\begin{document}
\begin{center}
{\Large Ring of physical states in the M(2,3) Minimal Liouville gravity}
\vspace{1mm}

%\vspace{1.0cm}
%\author{
%\centerline{A.~A.~Belavin${}^{a}$, V.~A.~Belavin${}^{b}$, A.~Litvinov${}^{a}$, Y.~Pugai${}^{a}$}
\vskip 1.0cm

{\large O.~Alekseev$^{a}$, M.~Bershtein$^{a,b}$ }

\vspace{.4cm} { \it
$^a$ Landau Institute for Theoretical Physics, 142432 Chernogolovka of Moscow Region, Russia\\

\vspace{0.3cm}
$^b$ Independent University of Moscow, 11 Bolshoy Vlasyevsky pereulok, 119002 Moscow, Russia\\
} \vspace{0.3cm} e-mail: alekseev@itp.ac.ru, mbersht@gmail.com

\vspace{1.0cm}
\textbf{Abstract}
\end{center}

We consider the $M(2,3)$ Minimal Liouville gravity, whose states in the
gravity sector are represented by irreducible modules of the Virasoro
algebra. We present a recursive construction for BRST cohomology classes.
This construction is based on using an explicit form of singular vectors in
irreducible modules of the Virasoro algebra. We construct an algebra of
operators acting on the BRST cohomology space. The operator algebra of
physical states is established by use of these operators.

%%%%%%%%%%%%%%%%%%%%%%%%%%%%%%%%%%%%%%%%%%%%%%%%%%%%%%%%%%%%%%%%%%%%%%%%%%%%%%%%%%%%%%%%%%%%%%%%%%%%%%%%%%%%%%
\section{Introduction} The Liouville gravity is a dynamic theory of the
metric on certain two-dimensional manifold whose action is induced by a
critical matter, i.e., matter described by a conformal field theory (CFT).
Simple reaction of conformal theories to the scaling of the metric leads to
the universal form of the effective action of the generated gravity called
Liouville gravity\cite{Polyakov}. In the David and Distler-Kawai (DDK)
approach\cite{DDK} the Liouville gravity can be represented as a tensor
product of a conformal matter theory, the Liouville theory, and a ghost
system. Schematically, the action for the Liouville gravity can be written
in the form
\begin{equation}
S=S^{M}+S^{L}+S^{gh}.\label{SMG}
\end{equation}
The consistency condition by David and Distler-Kawai impose a restriction on
the central charges of these theories. The restriction reads that the total
central charge of the critical matter, the Liouville theory and the ghosts
system vanishes
\begin{equation}
c_{L}+c_{M}+c_{gh}=0.\label{DDK}
\end{equation}
Thus in the critical gravity these three field theories are formally
decoupled and only interact due to the conformal anomaly cancellation
condition~(\ref{DDK}).

In this paper we consider the particular case of the Liouville gravity,
namely, a Minimal Liouville gravity where the conformal matter is a Minimal
CFT\cite{BPZ}. In this case it is possible to investigate a space of
physical states in detail.

The simplest states the ghost number\footnote{See section \ref{Notation} for
the definition of the ghost number.} 1 are the matter highest weight vectors
''dressed'' by an appropriate Liouville highest weight vectors such that the
total conformal dimension of the state (including ghosts) is equal to 0. The
simple structure of these states make it possible to study the corresponding
operators in detail. In particular, their three- and four-point functions
and an operator algebra have been found
explicitly~\cite{Zamolodchikov3poin,BelavinZamolodchikov}.

Lian and Zuckerman \cite{LianZuckerman} have realized that for a given
matter highest weight state there exists infinitely many additional states
with arbitrary ghost numbers. It is an interesting problem to consider
additional states in more detail and investigate the operator algebra of the
corresponding operators.

Note, that there are two versions of the Liouville gravity. In the first
version the gravity sector is realized by a theory of free scalar field.
From the mathematical point of view, it may be said that the space of states
in gravity sector is direct sum of the Feigin-Fuchs modules \cite{FF1}. In
this case the operator algebra has been investigated by Kanno and Sarmadi
\cite{KS}.

In the second version of the Liouville gravity the space of states in the
gravity sector is represented by irreducible modules \cite{ZamZam}. In has
been shown \cite{LianZuckerman} that these two versions of the Liouville
gravity possess different spaces of physical states.

In this paper we consider the second formulation of the Liouville gravity.
It is an interesting problem to consider the additional states. The natural
way to construct physical states is the BRST procedure. In the BRST
quantization procedure physical states are identified with BRST cohomology
classes. Generally, relative cohomology classes are called physical. In this
paper we present a recursive procedure to construct relative cohomology
classes. This generalizes construction of \cite{Imbimbo}

However we show that the definition of physical states as relative
cohomology classes is not completely suitable. The problem is that the
operator algebra of relative cohomology classes is not associative. In order
to avoid this problem we extend the space of states and consider absolute
cohomology classes. It is convenient to start from relative cohomology
classes and construct the absolute cohomology classes using certain
procedure.

We define certain operators that acts on the cohomology space. These
operators allow us to the calculate absolute cohomology and investigate the
operator algebra of absolute cohomology classes.

%%%%%%%%%%%%%%%%%%%%%%%%%%%%%%%%%%%%%%%%%%%%%%%%%%%%%%%%%%%%%%%%%%%%%%%%%%%%%%%%%%%%%%%%%%%%%%%%%%%%%%%%%%%%%%
\section{Notation}\label{Notation} The natural way to quantize the theory
(\ref{SMG}) with the constraint (\ref{DDK}) is the BRST procedure. Let
$(b,c)$ be the conformal ghost system of weights $(2,-1)$.
%\begin{equation}
%S_{gh}=\frac{1}{2\pi}\int d^2z(b\overline{\p}c+\overline{b}\p\overline{c}).
%\end{equation}
%The operator products are
%\begin{equation}
%b(z)c(0)\sim\frac{1}{z},\qquad b(z)b(0)\sim O(z),\qquad c(z)c(0)\sim O(z).
%\end{equation}
The ghost fields $b$ and $c$ admit the following Laurent expansions
$$ b(z)=\sum_{n=-\infty}^{\infty}\frac{b_n}{z^{n+2}},\qquad
c(z)=\sum_{n=-\infty}^{\infty}\frac{c_n}{z^{n-1}},$$ where the coefficients
$b_n$ and $c_n$ form an algebra with the only nonzero anticommutation
relation $$ \{b_m,c_n\}=\delta_{m+n,0}. $$
 We denote the Fock representation
of the ghost system by $\Lambda^{bc}$. Define a vacuum state $|0\ra_g$ in
$\Lambda^{bc}$ by the conditions $$ b_m|0\ra_g=0,~m\geq-1,\qquad
c_n|0\ra_g=0,~n\geq 2.$$ This vacuum is an $SL(2,C)$ invariant with the
conformal dimension~$0$. We assign a ghost number~$0$ to this vacuum. The
ghost field $b$ decreases the ghost number by 1, while $c$ increases the
ghost number by 1. It is convenient to define a vacuum
$|v^g\ra=c_1|0\ra_{g}$ of the conformal dimension~$-1$ and the ghost
number~$1$.

Let us consider a conformal field theory (CFT) with the central charge
$c=26$. In this paper we consider only chiral part of the CFT. Let $T(z)$
denote stress tensor. The modes $L_n$ of the stress tensor are given by the
Laurent expansion
$$
T(z)=\sum^\infty_{n=-\infty}{L_n\over z^{n+2}}.
$$
It was shown in \cite{BPZ} that operators $L_n$ satisfy commutation
relations
$$[L_n,L_m]=(n-m)L_{n+m}+\delta_{n+m,0}\frac{n^3-n}{12}c$$ and generate the
Virasoro algebra (Vir). Then the space of states in a conformal field theory
is representation of Vir.

 Let ${\cal M}$ be any representation of the CFT.
Introduce a Hilbert space
$$ C^{\rm abs}_*({\cal M})={\cal M}\otimes \Lambda^{bc}. $$
 Denote by $C^{\rm abs}_k({\cal M})$ the subspace of states of the definite ghost number~$k$.
The superscript `abs' stresses the fact that the space $C^{\rm abs}_*({\cal
M})$ forms an absolute BRST complex with respect to the BRST operator:
$$
\begin{aligned}
Q&=\oint:\lb T(z)+\frac{1}{2}T^{gh}(z)\rb c(z):-\frac{c_0}{2}=\\
&=\sum_{n=-\infty}^{\infty}
L_{-n}c_n-\frac{1}{2}\sum_{m,n=-\infty}^{\infty}(m-n):c_{-m}c_{-n}b_{n+m}:-\frac{c_0}{2},
\end{aligned}
$$
where $T^{gh}(z)$ is the stress tensor for the ghost system. It is well
known that $Q^2=0$ if and only if $c_{\cal M}=26$. We denote the cohomology
space of $C^{\rm abs}_*({\cal M})$ by $H^{\rm abs}_*({\cal M})$.

Let us introduce the subcomplex $$ C^{\rm rel}_*({\cal M})=\{w\in C^{\rm
abs}_*({\cal M})~|~b_0w=(L_0+L_0^{gh})w=0\}.$$
% where the states $w$ are restricted to be annihilated by $b_0$.
This subcomplex is called the relative BRST complex. We denote its
cohomology by $H^{\rm rel}_*({\cal M})$.

In this paper we restrict our consideration to the $M(2,3)$ Minimal gravity,
where the conformal matter is a Minimal CFT $M_{2,3}$ with the central
charge $c_M=0$. The only matter primary field possess the conformal
dimension $\Delta_M=0$ and the only representation in the Hilbert space is
the identity representation ${\mathbb I}$. One may say that the matter
sector of the model is trivial. The gravitational sector is represented by
the direct sum of the Virasoro irreducible modules ${\cal L}_{\Delta}$ with
the central charge $c_{L}=26$ and the highest weights $\Delta \in
\mathbb{C}$. Thus, the case
$$
{\cal M}=\bigoplus\limits_{\Delta\in\mathbb{C}}^{}\big( {\mathbb I}\otimes
{\cal L}_{\Delta} \big)=\bigoplus\limits_{\Delta\in\mathbb{C}}^{}{\cal
L}_{\Delta}
$$
corresponds to the Minimal Liouville Gravity $M(2,3)$.

Let $|{\cal L}_{\Delta}\ra $ be the highest weight vector in the Vir
irreducible module ${\cal L}_{\Delta}$. It is convenient to define a vacuum
vector in $C_*({\cal L}_{\Delta})$ by
\begin{equation}
\Psi_{\Delta}=|{\cal L}_{\Delta}\ra\otimes|v_g\ra,\label{Psi}
\end{equation}
In the BRST quantization procedure the physical states $w$ are defined to be
BRST cohomology classes, that is $Qw=0$ where the states $w$ are not BRST
exact.

%%%%%%%%%%%%%%%%%%%%%%%%%%%%%%%%%%%%%%%%%%%%%%%%%%%%%%%%%%%%%%%%%%%%%%%%%%%%%%%%%%%%%%%%%%%%%%%%%%%%%%%%%%%%%%
\section{Relative BRST complex}

%%%%%%%%%%%%%%%%%%%%%%%%%%%%%%%%%%%%%%%%%%%%%%%%%%%%%%%%%%%%%%%%%%%%%%%%%%%%%%%%%%%%%%%%%%%%%%%%%%%%%%%%%%%%%%
\subsection{Lian--Zuckerman theorems} \label{LZtheorems}

First, we describe Lian--Zuckerman theorem for irreducible Vir
representations. Let us consider the relative BRST complex $C_*^{\rm
rel}({\cal L}_{\Delta})$, where ${\cal L}_{\Delta}$ is the irreducible Vir
module with the highest weight $\Delta$. The cohomology $H_{*}^{\rm
rel}({\cal L}_\Delta)$ depend on the value of the highest weight~$\Delta$.
More precisely, $H_{*}^{\rm rel}({\cal L}_{\Delta})$ is non-trivial for
$\Delta$ belongs to some countable set $E=\{a_n, b_n\}$ of complex numbers.
This numbers are given in terms of the Kac conformal dimensions
$$
\Delta_{r,s}={25-(3r+2s)^2\over24}
$$
as \eq{ a_n=\Delta_{1,1+3(n-1)},\ n\geq0 \quad{\rm and}\quad
b_n=\Delta_{1,2+3(n-1)},\ n\geq1.\label{abdimensions} }

This numbers appear in the study of the structure of Verma modules with
$c=26$ \cite{FF2}. By $V_\Delta$ denote the Verma module with highest weight
$\Delta$ and central charge $c=26$. The Verma module $V_{a_o}$ is
irreducible: $V_{a_0}={\cal L}_{a_0}$. But the Verma modules $V_{a_1}$ and
$V_{b_1}$ possess a null vector of the weight $a_0$. It means that both
$V_{a_1}$ and $V_{b_1}$ contain the module $V_{a_0}$ as a submodule. This
can be continued: there is an infinite ladder of Verma modules $V_{a_k}$,
$V_{b_k}$, $k=1,2,\ldots$ that contain the modules $V_{a_{k-1}}$ and
$V_{b_{k-1}}$ as submodules. This can be represented by the following
embedding diagram \cite{FF2}: \nopagebreak
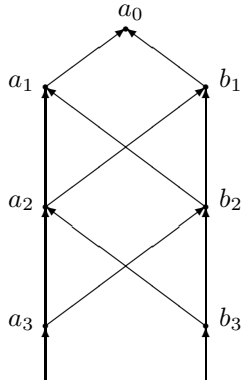
\begin{center}
\begin{picture}(90,140)
\put(50,132){\circle*{2}}
\put(47,137){$a_0$}

\put(20,110){\vector(4,3){30}}
\put(20,110){\circle*{2}}
\put(6,110){$a_1$}

\put(80,110){\vector(-4,3){30}}
\put(80,110){\circle*{2}}
\put(85,110){$b_1$}

\put(80,65){\vector(-4,3){60}}
\put(80,65){\vector(0,1){45}}
\put(80,65){\circle*{2}}
\put(85,65){$b_2$}

\put(20,65){\vector(4,3){60}}
\put(20,65){\vector(0,1){45}}
\put(20,65){\circle*{2}}
\put(6,65){$a_2$}

\put(80,20){\vector(-4,3){60}}
\put(80,20){\vector(0,1){45}}
\put(80,20){\circle*{2}}
\put(85,20){$b_3$}

\put(20,20){\vector(4,3){60}}
\put(20,20){\vector(0,1){45}}
\put(20,20){\circle*{2}}
\put(6,20){$a_3$}

\put(20,0){\vector(0,1){20}}
\put(80,0){\vector(0,1){20}}

\end{picture}
\end{center}
%\begin{figure}[htbp]\center
%\includegraphics[width=30mm]{embeddingdiagram.eps}\label{embeddingdiagram}
%\end{figure}

By the (embedding) level of the modules $V_{a_k}$, $V_{b_k}$ we shall call
the value $k$. An arrow connecting two nodes $\Delta\rightarrow \Delta'$
represents the fact that the module $V_{\Delta'}$ is a submodule of the
module~$V_{\Delta}$. In this case, the highest vector $|V_{\Delta'}\ra$ as a
vector in the module $V_\Delta$ is a null vector of the form
$D_{\Delta',\Delta}|V_\Delta\rangle$ with the operator $D_{\Delta',\Delta}$
being a linear combination of products of the Virasoro generators $L_{-k}$,
$k>0$.
%In this form the operator is defined uniquely up to a factor\cite{Zamolodchikov}.

Lian and Zuckerman proved in \cite{LianZuckerman} that the relative
cohomology classes $H^{\rm rel}_* ({\cal L}_{\Delta})$ are non-trivial if
and only if $\Delta \in E$. For each $\Delta\in E$ the dimension of the
cohomology space $H^{\rm rel}_*({\cal L}_{\Delta})$ is given by
\begin{eqnarray}
\dim H^{rel}_k({\cal L}_{a_n})=\dim H^{rel}_k({\cal L}_{b_n})=\left\{
\begin{aligned}
&1,\quad k=-n+1,~n+1, \\
&2,\quad k=-n+3,~-n+5,\ldots,~n-1,\\
&0, \quad \text{otherwise},
\end{aligned} \right. \label{LZLiouville}
\end{eqnarray}
where we assume that $n>0$. In the case $n=0$ the dimension of the
cohomology space is $$\dim H^{\rm rel}_k({\cal L}_{a_0})=\delta_{k,1}.$$

Second, we will need Lian--Zuckerman theorem for Verma modules. Consider the
BRST complex $C^{\rm rel}_*(V_{\Delta})$. In this case Lian and Zuckerman
also proved in \cite{LianZuckerman} that the relative cohomology space
$H^{\rm rel}_*(V_{\Delta})$ is non-trivial if and only if $\Delta \in E$.
The dimension of $H^{\rm rel}_*(V_{\Delta})$ is the following
\begin{equation}
 \dim H^{\rm rel}_k(V_{a_n})= \dim H^{\rm rel}_k(V_{b_n})=\left\{
\begin{aligned}
&1,\quad k=n+1, \\
&0, \quad \text{otherwise}.
\end{aligned} \right.\label{LZVerma}
\end{equation}

For any $\Delta$ there is a map $V_\Delta\rightarrow {\cal L}_\Delta$ from
Verma module with highest weight $\Delta$ to irreducible Vir module with the
same highest weight. It induces maps $C^{\rm rel}_*(V_{\Delta})\rightarrow
C^{\rm rel}_*({\cal L}_{\Delta})$ and $H^{\rm rel}_*(V_{\Delta})\rightarrow
H^{\rm rel}_*({\cal L}_{\Delta})$.
%Consequently, the only cohomology class from the space $H_*(V_{\Delta})$ has
%an image in the cohomology space $H_*({\cal L}_{\Delta})$.
Comparing the ghost numbers we see that the image of unique class in $H^{\rm
rel}_*(V_\Delta)$ is a cohomology class in $H^{\rm rel}_*({\cal
L}_{\Delta})$ of the highest ghost number. The cohomology class of the
highest ghost number in space $H^{\rm rel}_* ({\cal L}_{\Delta})$ will be
called the highest cohomology \footnote{ \label{correspon} As was already
mentioned in the introduction in another approach the space of states in
gravity sector is represented by free field (Feigin-Fuchs) modules. The
space of physical states in this theory is $H^{\rm rel}_*({\cal F})$, where
${\cal F}$ is a free field module with central charge $c_L=26$. In the case
$c \geq 25 $, the free field module is isomorphic to either a Verma module
or a contragradient Verma module (see \cite{Fr}). In the first case
cohomology $H^{\rm rel}_*({\cal F})=H^{\rm rel}_*(V_\Delta)$ and corresponds
to highest cohomology of $H^{\rm rel}_*({\cal L}_\Delta)$. One can show that
in second case cohomology $H^{\rm rel}_*({\cal F})$ corresponds to lowest
cohomology of $H^{\rm rel}_*({\cal L}_\Delta)$. This remark allows us to
compare our results with those of \cite{KS}, \cite{Govin1}, \cite{Govin2}}.

%%%%%%%%%%%%%%%%%%%%%%%%%%%%%%%%%%%%%%%%%%%%%%%%%%%%%%%%%%%%%%%%%%%%%%%%%%%%%%%%%%%%%%%%%%%%%%%%%%%%%%%%%%%%%%
\subsection{Recursive Construction of the Basis States}

We suggest that there exists a relation between expressions for physical
states and a form of the corresponding singular vectors. This relation leads
to an explicit recursive construction of cohomology classes. More precisely,
we show that all cohomology classes can be determined in terms of the
highest ones.

%%%%%%%%%%%%%%%%%%%%%%%%%%%%%%%%%%%%%%%%%%%%%%%%%%%%%%%%%%%%%%%%%%%%%%%%%%%%%%%%%%%%%%%%%%%%%%%%%%%%%%%%%%%%%%
\subsubsection{The Highest Cohomology Classes}

The construction of the highest cohomology classes simplifies due to
\begin{prop}
All the highest cohomology classes can be obtained by applying operators
$c_{-1},c_{-2},\ldots$ to the vacuum vectors $\Psi_{\Delta}$
intoduced~in~(\ref{Psi}).
\end{prop}
This proposition easily follows from Proposition~1.11 in~\cite{FGZ}.
An~essential part of the proof of this proposition is the following
construction of the highest cohomology classes. Let $K^n$ be the vector
space of~all (possibly~infinite) linear combinations of antisymmetric
monomials
\begin{equation}
c_{-i_1,\ldots, -i_n}=c_{-i_1}c_{-i_2}\cdots c_{-i_n}.\label{ci1...in}
\end{equation}
Let $d(c_{-i_1,\ldots, -i_n})=i_1+i_2+\cdots+i_n$ be the degree of the
monomial (\ref{ci1...in}). Let us define a differential $\delta : K^n
\rightarrow K^{n+1} $ as follows
\begin{equation}
\delta(c_{-i})=\sum\limits_{\alpha+\beta=i}
(\alpha-\beta)c_{-\alpha,-\beta}, \qquad \delta(c_{-i_1,\ldots,-i_n})=\sum\limits_{j=1}^{n} (-1)^{j-1} d(c_{-i_j})
c_{-i_1,\ldots,-\hat{i}_j,\ldots, -i_n}.\nn
\end{equation}
It is easily to prove that $\delta^2=0$. We denote the cohomology space of
this complex by $H^*(K)$. This complex is isomorphic to the standard
cohomology complex of the Lie algebra $Vir_{>0}=\la L_1,L_2,\ldots \ra$ and
was studied well. The Goncharova theorem \cite{Goncharova} states that the
dimensions of the homology spaces $H^n(K)$ is
\begin{equation}
\dim H^n(K)=\left\{
\begin{aligned}
&1, \qquad n=0,\\
&2,\qquad n>0.
\end{aligned}\right.\nn
\end{equation}
Each cohomology space $H^n(K)$ is generated by two vectors $u_n,v_n$ with
the degrees
\begin{equation}
d(u_n)=\frac{3n^2-n}{2},\qquad d(v_n)=\frac{3n^2+n}{2}.\label{dun}
\end{equation}
 For example,
$$H^1(K)=\la c_{-1}, c_{-2} \ra\quad{\rm and}\quad H^2(K)=\la c_{-1}c_{-4},
c_{-2}c_{-5}-3c_{-3}c_{-4} \ra.
$$
The vectors $u_n$ and $v_n$ are defined modulo $\delta$ exact terms.

 Let us return to the construction of the highest cohomology classes for the Vir
irreducible modules ${\cal L}_{a_n}$. Consider the vector
$$
u_n\Psi_{a_n}=|{\cal L}_{a_n}\ra\otimes u_n |v^g\ra
$$
of the ghost number $n+1$ and the conformal dimension 0 (since
$a_n+d(u_n)-1=0$ where $a_n$ in given in~(\ref{abdimensions}) and $d(u_n)$
is given in~(\ref{dun})). It is clear that the vector $u_n\Psi_{a_n}$ is
BRST closed, i.e.,~$Q(u_n\Psi_{a_n})=0$. Moreover, one can show that this
vector is not BRST exact. Thus, the state
\begin{equation}
O_{a_n}^{a_n}=u_n\Psi_{a_n}.\nn
\end{equation}
is a representative of the relative cohomology $H^{\rm rel}_{n+1}({\cal
L}_{a_n})$.

The highest cohomology classes of any Vir irreducible module ${\cal
L}_{b_n}$ can be constructed in a similar manner. We give an explicit form
of several highest cohomology classes
\begin{equation}
\begin{aligned}
O_{a_1}^{a_1}&=H_{a_1}^{a_1}\Psi_{a_1}=c_{-1}\Psi_{a_1}, &O_{a_2}^{a_2}&=H_{a_2}^{a_2}\Psi_{a_2}=c_{-1}c_{-4}\Psi_{a_2},\\
O_{b_1}^{b_1}&=H_{b_1}^{b_1}\Psi_{b_1}=c_{-2}\Psi_{b_1},
&O_{b_2}^{b_2}&=H_{b_2}^{b_2}\Psi_{b_2}=(c_{-2}c_{-5}-3c_{-3}c_{-4})\Psi_{b_2}.
\end{aligned}
\label{HighestCo}
\end{equation}

%%%%%%%%%%%%%%%%%%%%%%%%%%%%%%%%%%%%%%%%%%%%%%%%%%%%%%%%%%%%%%%%%%%%%%%%%%%%%%%%%%%%%%%%%%%%%%%%%%%%%%%%%%%%%%
\subsubsection{Recursion equations}

In this subsection we use the following notation. Let us introduce the BRST
complex~$C_*(V_{\Delta})$, where~$V_{\Delta}$ is the Verma module with
highest weight~$\Delta$. Let $|V_{\Delta}\ra$ be the highest weight vector
in this Verma module. We define a~vacuum vector in this complex by
\begin{equation}
\Psi^V_{\Delta}=|V_{\Delta}\ra\otimes| v^g \ra.\nn
\end{equation}
To emphasize the difference between this vacuum vector and the vacuum vector
defined in~(\ref{Psi}) we denote the latter by $\Psi^{\cal L}_{\Delta}$.

The cohomology classes, except the highest ones, possess recursive
construction\footnote{ The meaning of this construction is the following. We
take the resolution of a Vir irreducible module by Verma modules \cite{FF2}
and calculate the BRST cohomology with coefficients in this resolution. Then
the cohomology classes for Vir irreducible modules can be obtained by
spectral sequence.}. This construction makes it possible to find the
expressions for cohomology classes related to the~highest weights of the
Verma modules in the embedding diagram one by one downwards starting from
the top. Let us perform several first steps explicitly. Then we shall
describe the~$n$-th~step.

\textbf{Embedding level 0.}\nopagebreak

The top node of the embedding diagram corresponds to the highest weight
$a_0=1$. The dimension of the cohomology space is $\dim H^{\rm rel}_k({\cal
L}_{a_0})=\delta_{k,1}$. It is straightforward to find a representative of
cohomology space. But for further references we give an additional
construction. Let us consider the relative BRST complex $C^{\rm
rel}_*(V_{a_0})$. It is easy to check that the state \eq{
H_{a_0}^{a_0}\Psi^V_{a_0}=\Psi^V_{a_0}\label{PrimaryStateVir} } is a
representative of the cohomology classes. Taking into account our discussion
in the last part of Subsection \ref{LZtheorems}, one can check that the
state

\eq{ O_{a_0}^{a_0}=H_{a_0}^{a_0}\Psi^{\cal L}_{a_0}=\Psi^{\cal
L}_{a_0}.\label{PrimaryState} } is a representative of the cohomology
classes $H^{\rm rel}({\cal L}_{a_0})$. This state is the simplest and was
the subject of the most of studies of correlation functions. The
corresponding operator contains a primary field in the matter sector (in our
case it is an identity operator) dressed by an appropriate Liouville primary
field.

\textbf{Embedding level 1.}\nopagebreak

Nodes at the first level of the embedding diagram corresponds to the highest
weights $a_1=0$ and $b_1=-1$. Let us consider the cohomology space $H^{\rm
rel}_*({\cal L}_{a_1})$ associated with the former. By Lian--Zuckerman
results (\ref{LZLiouville}), one can realize that non-trivial cohomology
classes belong to $H^{\rm rel}_0({\cal L}_{a_1})$ and $H^{\rm rel}_2({\cal
L}_{a_1})$. The highest cohomology class $O_{a_1}^{a_1} \in H^{\rm
rel}_2({\cal L}_{a_1})$ is given in (\ref{HighestCo}). The cohomology class
from the space $H^{\rm rel}_0({\cal L}_{a_1})$ can be constructed as
follows:

{\bfseries\itshape Step 1.} Let us consider the relative BRST complex
$C^{\rm rel}_*(V_{a_1})$. The Verma module $V_{a_1}$ contains a singular
vector at the first level. This singular vector has the form
$D_{a_0,a_1}|V_{a_1}\ra$ for some operator $D_{a_0,a_1}$. Here we have
$D_{a_0,a_1}=L_{-1}$. Let us define the state

\eq{ O_{a_0|a_1}^{a_0}=H^{a_0}_{a_0}D_{a_0,a_1}\Psi^V_{a_1} \in C^{\rm
rel}_*(V_{a_1}).\label{Oa1a0tilde} }

Note that the Verma modules with the highest weight vectors
$D_{a_0,a_1}|V_{a_1}\ra$ and $|V_{a_0}\ra$ are equivalent, since both
modules have the same highest weight $a_0=1$. Thus, one can consider the
state (\ref{Oa1a0tilde}) as $H_{a_0}^{a_0}\Psi^V_{a_0}$ determined on the
previous level (\ref{PrimaryStateVir}). Consequently, the state
(\ref{Oa1a0tilde}) is BRST closed, since the state (\ref{PrimaryStateVir})
is BRST closed.

{\bfseries\itshape Step 2.} We will show that the state (\ref{Oa1a0tilde})
is also BRST exact. Indeed, due to Lian-Zuckerman results~(\ref{LZVerma})
the only non-trivial cohomology class in the space $H^{\rm
rel}_{*}(V_{a_1})$ has the ghost number~2, while the
state~(\ref{Oa1a0tilde}) has the ghost number~1. Therefore, this state is
BRST exact. Hence, there exists a certain operator $H_{a_1}^{a_0}$ such that
\begin{equation}
Q(H_{a_1}^{a_0}\Psi^V_{a_1})=O_{a_0|a_1}^{a_0}.\label{D11Va1}
\end{equation}
%We shall call this equation as the recursive equation since the right side
%of this equation is completely specified by unique cohomology class at the
%previous level and the embedding structure of the Verma modules.
The right side of this equation is completely specified by the unique
cohomology class at the previous level and the embedding structure of the
Verma modules. Therefore, in the sequel this equation will be called the
recursive equation. A solution of the equation~(\ref{D11Va1}), namely, the
operator $H_{a_1}^{a_0}$ is used to obtain a representative of the
cohomology space $H^{\rm rel}_0({\cal L}_{a_1})$ in the following:

{\bfseries\itshape Step 3.} We assert that the state
$O_{a_1}^{a_0}=H_{a_1}^{a_0}\Psi^{\cal L}_{a_1}$ represents a cohomology
class in $H^{\rm rel}_0({\cal L}_{a_0})$. Indeed by~(\ref{D11Va1})
and~(\ref{Oa1a0tilde}) we get\footnote{There is a map $V_{a_1}\rightarrow
{\cal L}_{a_1}$ from the Verma module to the irreducible Vir module.
Consequently, there exists the map $H_*^{\rm rel}(V_{a_1})\rightarrow H^{\rm
rel}_*({\cal L}_{a_1})$. The r.h.s. of (\ref{D11Va1}) is in the kernel of
this map, while $H_{a_1}^{a_0}\Psi^V_{a_1}$ has an image
$H_{a_1}^{a_0}\Psi^L_{a_1}$, which we denoted by $O_{a_1}^{a_0}$. Under this
map the equation (\ref{D11Va1}) takes the form $Q(O_{a_1}^{a_0})=0$.}
$$
Q(O_{a_1}^{a_0})=H^{a_0}_{a_0}D_{a_0,a_1}|{\cal L}_{a_1}\ra\otimes
|v^g\ra=0,$$ since $D_{a_0,a_1}|{\cal L}_{a_1}\ra=0$. Moreover, one can show
that the state~$O_{a_1}^{a_0}$ is not BRST exact. Thus it represents a
cohomology class.

We have obtained all cohomology classes in the space $H^{\rm rel}({\cal
L}_{a_1})$ and show the connection between these classes and singular
vectors in the irreducible Vir module~${\cal L}_{a_1}$.

The cohomology classes in the space $H^{\rm rel}_{*}({\cal L}_{b_1})$ can be
obtained in a similar manner. The highest cohomology class is given in
(\ref{HighestCo}). The other part of the construction is the same as we used
in the space $H^{\rm rel}_0({\cal L}_{a_1})$. Let us list the results. There
is a singular vector at the second level in the Verma module $V_{b_1}$. This
vector has the form $D_{a_0,b_1}|V_{b_1}\ra$ for some operator
$D_{a_0,b_1}$. Here we have $D_{a_0,b_1}=(L_{-1}^2+(2/3)L_{-2})$. The
recursive equation reads
\begin{equation}
Q(H_{b_1}^{a_0}\Psi^V_{b_1})=O_{a_0|b_1}^{a_0},\qquad
O_{a_0|b_1}^{a_0}=H_{a_0}^{a_0}D_{a_0,b_1}\Psi^V_{b_1},\label{D12Vb1}
\end{equation}
The solution of this equation allows us to specify the operator
$H_{b_1}^{a_0}$ and, hence, the representative
$O_{b_1}^{a_0}=H_{b_1}^{a_0}\Psi^{\cal L}_{b_1}$ of the cohomology space
$H^{\rm rel}_{0}({\cal L}_{b_1})$.

\textbf{Embedding level 2}\nopagebreak

Nodes at this level of the embedding diagram corresponds to the highest
weights $a_2=-4$ and $b_2=-6$. We consider the cohomology space $H^{\rm
rel}({\cal L}_{a_2})$. The dimensions and ghost numbers of the cohomology
classes are given by Lian--Zuckerman results~(\ref{LZLiouville}). The
highest cohomology class $O_{a_2}^{a_2}$ is given in~(\ref{HighestCo}).

Let us consider the cohomology space $H_1^{\rm rel}({\cal L}_{a_2})$. Its
two basic cohomology classes can be constructed in a similar manner as on
the previous level. There are two singular vectors in the Verma
module~$V_{a_2}$. The first one $D_{a_1,a_2}|V_{a_2}\ra$ is at the level~4
and the second one $D_{b_1,a_2}|V_{a_2}\ra$ is on the level~3. In this case
the recursive equations are

\eq{ \Aligned{
&Q(H_{a_2}^{a_1}\Psi_{a_2}^V)=O_{a_1|a_2}^{a_1},\qquad &O_{a_1|a_2}^{a_1}=H_{a_1}^{a_1}D_{a_1,a_2}\Psi^V_{a_2},\\
&Q(H_{a_2}^{b_1}\Psi_{a_2}^V)=O_{b_1|a_2}^{b_1},\qquad
&O_{b_1|a_2}^{b_1}=H_{b_1}^{b_1}D_{b_1,a_2}\Psi^V_{a_2}.\label{level2recEq}
} }

The operators $H_{a_2}^{a_1}$ and $H_{a_2}^{b_1}$ specify the
representatives $O_{a_2}^{a_1}=H_{a_2}^{a_1}\Psi^{\cal L}_{a_2}$ and
$O_{a_2}^{b_1}=H_{a_2}^{b_1}\Psi^{\cal L}_{a_2}$ of the cohomology classes
$H_{1}^{\rm rel}({\cal L}_{a_2})$.

The construction of the representative of the cohomology space $H^{\rm
rel}_{-1}({\cal L}_{a_2})$ is a little more tricky.

{\bfseries\itshape Step 1.} Let us consider the relative BRST complex
$C^{\rm rel}_*(V_{a_2})$. As we discussed, there are two singular vectors in
the Verma module $V_{a_2}$. Let us introduce the states
$$
O_{a_1|a_2}^{a_0}=H_{a_1}^{a_0}D_{a_1,a_2}\Psi^V_{a_2},\quad
O_{b_1|a_2}^{a_0}=H_{b_1}^{a_0}D_{b_1,a_2}\Psi_{a_2}^V.
$$
One can consider these states as $H_{a_1}^{a_0}\Psi^V_{a_1}$ and
$H_{b_1}^{a_0}\Psi^V_{b_1}$, which are specified at the previous level.
Thus, by relations (\ref{Oa1a0tilde}), (\ref{D11Va1}) and (\ref{D12Vb1}) we
obtain

\eq{
Q(O_{a_1|a_2}^{a_0})=H_{a_1}^{a_0}D_{a_0,a_1}D_{a_1,a_2}\Psi^V_{a_2},\quad
Q(O_{b_1|a_2}^{a_0})=H_{b_1}^{a_0}D_{a_0,b_1}D_{b_1,a_2}\Psi^V_{a_1}.\label{Q2level2}
}

As we discussed above, there are a singular vectors of conformal dimension
$a_0=1$ in the Verma modules $V_{a_1}$ and $V_{b_1}$. This modules and,
therefore, the singular vectors are contained in the Verma module $V_{a_2}$.
Since the Verma module $V_{a_2}$ contains the only singular vector of the
highest weight~$a_0$, one get an operator identity
$D_{a_0,a_1}D_{a_1,a_2}=D_{a_0,b_1}D_{b_1,a_2}$. Thus, from~(\ref{Q2level2})
we obtain
$$ Q(O_{a_1|a_2}^{a_0}-O_{b_1|a_2}^{a_0})=0. $$
The remaining part of the construction is the same as we used on the
previous level. Thus, let us list the results. The recursive equation is
\begin{equation}
Q(H_{a_2}^{a_0}\Psi^V_{a_2})=O_{a_1|a_2}^{a_0}-O_{b_1|a_2}^{a_0},\label{Ha2a0}
\end{equation}
and $O^{a_2}_{a_0}=H_{a_2}^{a_0}\Psi^{\cal L}_{a_2}$ is a cohomology class
in the space $H^{\rm rel}_{-1}({\cal L}_{a_2})$. As we can see the equation
(\ref{Ha2a0}) determine the cohomology class at the second level by
cohomology classes at the first level.

\textbf{Embedding level n}\nopagebreak

We suggest that one can obtain the recursive equations for cohomology
classes of the space $H_*({\cal L}_{a_n})$. The basic cohomology classes and
their ghost numbers $N^g$ can be represented in the form of the following
diagram: \nopagebreak
\begin{center}
\begin{picture}(300,110)
\put(40,70){$O_{a_n}^{a_0}$} \put(90,40){$O_{a_n}^{b_{1}}$}
\put(90,100){$O_{a_n}^{a_{1}}$} \put(140,40){$O_{a_n}^{b_{2}}$}
\put(140,100){$O_{a_n}^{a_{2}}$} \put(190,40){$\ldots$}
\put(190,100){$\ldots$} \put(240,40){$O_{a_n}^{b_{n-1}}$}
\put(240,100){$O_{a_n}^{a_{n-1}}$} \put(290,80){$O_{a_n}^{a_n}$}
\put(0,0){$N^{g}=$} \put(30,0){{ $-n+1,$}} \put(80,0){{ $-n+3,$}}
\put(130,0){{ $-n+5,$}} \put(187,0){$\ldots$} \put(240,0){{ $n-1,$}}
\put(287,0){{ $n+1.$}}
\end{picture}
\end{center}

%\begin{figure}[htbp]\center
%\includegraphics[width=80mm]{basis.eps}\label{basisRef}
%%\caption{The basis in the space $H_*^{rel}{\cal L}_{a_k}$}
%\end{figure}

Let $\gamma_k$ and $\delta_k$ be either $a_k$ or $b_k$. Define the set of
operators $H_{a_n}^{\gamma_k}$ by $$
O_{a_n}^{\gamma_k}=H_{a_n}^{\gamma_k}\Psi^{\cal L}_{a_n}.$$ These operators
are polynomials in Vir algebra generators $L_{n}$ and ghosts $c_n$, $b_n$
with $n<0$. Let us introduce the set of states
$$
O_{\delta_{n-1}|a_n}^{\gamma_{j}}=H_{\delta_{n-1}}^{\gamma_{j}}D_{\delta_{n-1},\
a_n}\Psi^V_{a_n}.
$$
One can consider these state as the states
$H_{\delta_{n-1}}^{\gamma_{j}}\Psi^V_{\delta_{n-1}}$,~i.e., as a cohomology
classes on the previous level $n-1$. Thus, by the assumption of the
recursion procedure the set of the operators $H_{\delta_{n-1}}^{\gamma_{j}}$
is supposed to be specified. The operators $D_{\delta_{n-1},\ a_n}$ are
specified by the embedding structure of singular vectors in the Verma module
$V_{a_n}$. Now we can formulate
\begin{prop}\label{RelBasisProp}
For the cohomology classes of the space $H^{\rm rel}_*({\cal L}_{a_n})$ the
set of recursive equations reads
\begin{equation}
Q(H_{a_n}^{\gamma_{j}}\Psi_{a_n}^V)=\left\{
\begin{aligned}
&O_{a_{n-1}|a_n}^{\gamma_{j}}-O_{b_{j}|a_n}^{\gamma_{j}},\quad &&j=0,\ldots, n-2\\
&O_{\gamma_{n-1}|a_n}^{\gamma_{n-1}},\quad&&j=n-1\\
&0,\quad &&j=n,
\end{aligned} \right.\nn
\end{equation}
and states $O_{a_n}^{\gamma_{j}}=H_{a_n}^{\gamma_{j}}\Psi_{a_n}^{\cal L}$
form a basis in the cohomology space $H^{\rm rel}_*({\cal L}_{a_n})$.
\end{prop}
One can formulate a similar proposition for the cohomology space $H^{\rm
rel}_*({\cal L}_{b_n})$. As a consequence of the Proposition
\ref{RelBasisProp}, we show that forms for all cohomology classes can be
determined from the highest ones and the operators $D_{\Delta',\Delta}$. The
explicit expressions for the operators $D_{\Delta',\Delta}$ was found in
\cite{BdFIZ}.

%%%%%%%%%%%%%%%%%%%%%%%%%%%%%%%%%%%%%%%%%%%%%%%%%%%%%%%%%%%%%%%%%%%%%%%%%%%%%%%%%%%%%%%%%%%%%%%%%%%%%%%%%%%%%%

%%%%%%%%%%%%%%%%%%%%%%%%%%%%%%%%%%%%%%%%%%%%%%%%%%%%%%%%%%%%%%%%%%%%%%%%%%%%%%%%%%%%%%%%%%%%%%%%%%%%%%%%%%%%%%

\subsection{Operators Acting on the Relative Cohomology Space}

In the previous subsection we constructed the basis in the cohomology space
$H^{\rm rel}_*({\cal L}_{a_n})$. To investigate the operator algebra it is
convenient to choose another set of representatives of the cohomology
classes. We construct a new basis by introducing certain operators acting on
the cohomology space.

Let us consider two operators
\begin{equation}
X=\frac12\sum_{n=-\infty}^{\infty} n c_{-n} c_n,\qquad
X_+=\frac12\sum_{n=-\infty}^{\infty} n^3 c_{-n} c_n.\nn
\end{equation}
These operators commute with the BRST charge $Q$. Indeed, it is easy to
check that the operator $X$ is equal to $[Q,c_0]$ and commutes with the BRST
charge due to the property $Q^2=0$. The commutation of the operator $X_+$
with the BRST charge can be verified straightforwardly. Thus these operators
act on the relative cohomology space, i.e., if $w$ represents a BRST
cohomology class of the ghost number~$k$, then $Xw$ and $X_+w$ represent
cohomology classes of the ghost numbers~$k+2$ in the same space.

 The operators $X$ and $X_+$ satisfy the relation
$$ X\cdot X_+=0$$
on the cohomology space $H^{\rm rel}_*({\cal L}_{\Delta})$. This relation
can be verified as follows. Let us introduce the operator
\begin{equation}
\tilde{Y}=\frac1{12}\sum\limits_{\scriptstyle i+j+k=0,\atop\scriptstyle
i,j,k\neq0}(i-j)(j-k)(k-i)c_ic_jc_k.\label{tildeY}\end{equation} It is
straightforward to check that the product $X\cdot X_+$ is equal to
$[\tilde{Y},Q]$. Therefore, for any cohomology class $w$, we have
$XX_+w=[\tilde{Y},Q]w=-Q\tilde{Y}(w)$, i.e. $XX_+w$ is equal to 0 in the
cohomology space.

The operators $X$ and $X_+$ with relation $XX_+=0$ generate an algebra
acting on the cohomology space $H^{\rm rel}_*({\cal L}_{\Delta})$
\footnote{From more abstract point of view, we study an action of the
cohomology of Virasoro algebra on semi-infinite cohomology. The operators
$X$ and $X_+$ form a basis in the two-dimensional space
$H^2(Vir,Vir_0,\mathbb{C})$, where $Vir_0=\la L_0,c\ra$. The whole algebra
$H^*(Vir,Vir_0,\mathbb{C})$ is generated by $X$ and $X_+$ with one relation
$XX_+=0$. }.

Operators $X$ and $X_+$ can be used to construct representatives of the
cohomology classes. We construct a representatives of the cohomology classes
by use of this operators. Consider the space $H^{\rm rel}_{*}({\cal
L}_{a_n})$. To simplify notation denote by $O_{a_n}$ the cohomology class
$O^{a_0}_{a_n}$ of the lowest ghost number. A basis in the cohomology space
can be obtained by means of
\begin{thm}\label{newbase} Cohomology classes
\begin{equation}
\begin{aligned}
O_{a_n}, \qquad XO_{a_n},\ X^2O_{a_n},\ldots X^nO_{a_n},\qquad X_+O_{a_n},\
X^2_+O_{a_n},\ldots X^{n-1}_+O_{a_n}\nn
\end{aligned}
\end{equation}
form a basis in the cohomology space $H^{\rm rel}_* ({\cal L}_{a_n})$.
\end{thm}

The Theorem \ref{newbase} can be illustrated as follows \nopagebreak
\begin{center}
\begin{picture}(380,120)
\put(40,70){$O_{a_n}$} \put(55,75){\vector(4,3){35}}
\put(55,75){\vector(4,-3){35}} \put(65,90){$X$} \put(65,50){$X_+$}

\put(95,100){$XO_{a_n}$} \put(95,40){$X_+O_{a_n}$}

\put(120,104){\vector(1,0){30}} \put(120,44){\vector(1,0){30}}

\put(128,107){$X$} \put(128,47){$X_+$}

\put(155,100){$X^2O_{a_n}$} \put(155,40){$X^2_+O_{a_n}$}

\put(185,104){\vector(1,0){30}} \put(185,44){\vector(1,0){30}}

\put(193,107){$X$} \put(193,47){$X_+$}

\put(220,40){$\ldots$} \put(220,100){$\ldots$}

\put(235,104){\vector(1,0){30}} \put(235,44){\vector(1,0){30}}

\put(243,107){$X$} \put(243,47){$X_+$}

\put(270,40){$X^{n-1}_+O_{a_n}$} \put(270,100){$X^{n-1}O_{a_n}$}

\put(310,102){\vector(4,-3){37}} \put(310,46){\vector(4,3){37}}

\put(330,90){$X$} \put(330,50){$X_+$}

\put(350,70){$X^{n}O_{a_n}$}

\put(0,0){$N^g=$} \put(35,0){$-n+1,$} \put(90,0){$-n+3,$}
\put(150,0){$-n+5,$} \put(220,0){$\ldots$} \put(280,0){$n-1,$}
\put(350,0){$n+1.$}
\end{picture}
\end{center}

The similar theorem holds for the cohomology space $H^{\rm rel}_*({\cal
L}_{b_n})$. Proof of this theorems will be published elsewhere
\footnote{\label{duality} The idea of the proof is the following (this idea
is due to B.~Feigin) The algebra generated by $X,X_+$ with relation $XX_+=0$
is the cohomology algebra $H^*(Vir,Vir_0,\mathbb{C})$. The homology space
$H_*(Vir,Vir_0,\mathbb{C})$ is a cofree module over this algebra. The
irreducible Vir module $\mathbb{C}$ is dual to some infinite complex
$\mathcal{K}$ by means of duality between $c=0$ and $c=26$ Virasoro modules.
Hence the $H^{\infty/2+*}(Vir,Vir_0,\mathcal{K})$ is free module over the
algebra $H^*(Vir,Vir_0,\mathbb{C})$. The irreducible Liouville module ${\cal
L}_{a_n}$ is quasi-isomorphic to the truncation of the complex
$\mathcal{K}$. Therefore the space $H^{\rm rel}_* ({\cal
L}_{a_n})=H^{\infty/2+*}(Vir,Vir_0,{\cal L}_{a_n})$ is cyclic module over
algebra $H^*(Vir,Vir_0,\mathbb{C})$}. Let us check the Theorem \ref{newbase}
in the first nontrivial example $H^{\rm rel}_*({\cal L}_{a_2})$. By explicit
form of the cohomology classes (see Appendix) we obtain
\begin{equation}
\begin{aligned}
&X(O_{a_2}^{a_0})=-\frac{5}{3}\cdot O_{a_2}^{b_1}-O_{a_2}^{a_1},\qquad
\qquad&& X_+(O_{a_2}^{a_0})=\frac{7}{3}\cdot O_{a_2}^{b_1}-O_{a_2}^{a_1},
\\ & X^2(O_{a_2}^{a_0})=240 O_{a_2}^{a_2} \qquad \qquad && X^2_+(O_{a_2}^{a_0})=-3696
O_{a_2}^{a_2} \label{XX+a2}
\end{aligned}
\end{equation}
This implies that the cohomology classes $O_{a_2}, XO_{a_2}, X^2O_{a_2},
X_+O_{a_2}$ form a basis in the cohomology space $H^{\rm rel}_*({\cal
L}_{a_2})$. Also it follows from (\ref{XX+a2}) that the basis introduced in
Theorem~\ref{newbase} differs from that of Proposition~\ref{RelBasisProp}.

From Theorem~\ref{newbase} follows that $O_{a_n}^{a_n}=\lambda
X^nO_{a_0}^{a_n}$, where $\lambda \neq 0$, i.e. the highest cohomology class
can be obtained from lowest one by use of operator $X$. This fact concerns
only highest and lowest cohomology classes therefore it should have analogue
in free field approach due to remark~\ref{correspon}. Indeed this formula is
equivalent to the equation~(2.12)~in~\cite{Govin2}.

%\begin{figure}[htbp]\center
%\includegraphics[width=90mm]{operatorsRel.eps}\label{basis2}
%\end{figure}

%An arrow connecting two classes $X^kO_{a_n}\rightarrow X^{k+1}O_{a_n}$ means
%that the cohomology class $X^{k+1}O_{a_n}$ can be obtained from $X^kO_{a_n}$
%by applying the operator $X$. The same holds for the operator $X_+$.

Let us recall that all cohomology classes except highest ones can be
obtained from the previous level cohomology classes by the recursive
construction. On the other hand, it follows from the Theorem~\ref{newbase}
that all the cohomology classes (including the highest one) in the space
$H^{\rm rel}_*({\cal L}_{a_n})$ can be obtained from the cohomology class
with the lowest ghost number $O_{a_n}$ by applying the operators $X,X_+$.
Therefore all cohomology classes can be found from the simplest cohomology
$O_{a_0}^{a_0}\in H^{\rm rel}_*({\cal L}_{a_0})$ (see \ref{PrimaryState}).

\subsection{Operator Algebra}

One can construct physical operators from the states in the Hilbert space
using state-operator correspondence. Every state in the Hilbert space has an
image in the space of local operators. For example, any Liouville highest
weight vector corresponds to a Liouville primary field of the same conformal
dimension.

For any cohomology class $O$ we can consider a unique local operator $O(z)$,
commuting with the BRST charge $Q$. This operator doesn't depend on point
$z$ modulo BRST exact terms. Indeed,

$$\partial O(z)=L_{-1}O(z) \quad \quad  L_{-1}O=[Q,b_{-1}]O=Qb_{-1}O.$$
It is well known that any nontrivial BRST cohomology class O has zero
conformal dimension. The prove is simple. Suppose that $L_0O=\Delta O$ and
$\Delta \neq 0$. Hence, we have
$$O=\Delta^{-1} L_0O=\Delta^{-1} [Q,b_0]O= Q\left(\Delta^{-1} b_0 O\right) .$$

The general form of the operator product expansion (OPE) of any two local
operators $O_1(z)$ and $O_2(0)$ reads
\begin{equation}
O_1(z)O_2(0)=\sum_{n=-\infty}^{\infty} A_n(0)z^n,\nn
\end{equation}
where the expansion coefficients $A_n(0)$ are some operators of the
conformal dimension $n$. Since the operators the $O_1(z)$ and $O_2(0)$
commute with BRST charge Q, the expansion coefficients $A_n(z)$ commute with
it as well. Since there is no BRST cohomology classes of non-zero conformal
dimension, only $A_0(z)$ may give a BRST nontrivial cohomology class. Let us
denote $A_0(z)$ by $O_{3}(z)$. We obtain a ring structure, defined by the
operator product expansion modulo BRST exact terms
\begin{equation}
O_1(z)O_2(0)=O_{3}(0)+[Q,\ldots],\nn
\end{equation}
which will be denoted by
\begin{equation}
O_1\cdot O_2=O_{3}.\label{Ring}
\end{equation}

We assert that the operator algebra on the relative cohomology space is not
associative. Let us consider the simplest non-trivial example and show that
\begin{equation}
(O^{a_1}_{a_1}\cdot O^{a_1}_{a_1})\cdot O^{a_0}_{a_{2}}\neq
O^{a_1}_{a_1}\cdot (O^{a_1}_{a_1}\cdot O^{a_0}_{a_{2}}).\label{OOO}
\end{equation}

The l.h.s of (\ref{OOO}) is equal to~0 in the cohomology space. Indeed, from
Liouville fusion rules and ghost number conservation follows that
$O_{a_1}^{a_1}\cdot O_{a_1}^{a_1}$ belongs to $H^{\rm rel}_{*}({\cal
L}_{a_1})$ and has ghost number~$4$. Due to Lian-Zuckerman
results~(\ref{LZLiouville}) $H^{\rm rel}_{4}({\cal L}_{a_1})=0$. Therefore
$O_{a_1}^{a_1}\cdot O_{a_1}^{a_1}$ is equal to~0, then l.h.s of (\ref{OOO})
is equal to~0.

The r.h.s. of (\ref{OOO}) can be evaluated using the explicit form of the
operators given in Appendix. The r.h.s can be shown to be equal to $240
O_{a_2}^{a_2}$. This calculation proves that the operator algebra of
relative cohomology classes is not associative.

The absence of the associativity of the operator algebra is quite
undesirable. Let us investigate this problem in more detail. Up to now we
discussed the relative BRST cohomology classes $w$, modulo $Qw'$ where both
elements $w$ and $w'$ are annihilated by $b_0$. Note that there exist states
of the form $Q\tilde{w}$ such that $b_0\tilde{w}\neq 0$. For example, the
state $O_{a_1}^{a_1}$ is of this kind
$$
O_{a_1}^{a_1}=Q\left(c_0b_{-1}\Psi_{a_1}\right).$$ Any correlation function
that contains the such states vanishes. One can say that these states are
not physical.

Thus we need to exclude such states from the physical spectrum. In order to
exclude undesirable states we consider the absolute BRST complex in the next
section.

\section{Absolute cohomology}

\subsection{Basic cohomology classes} In this section we consider the
absolute BRST complex $C^{\rm abs}_*({\cal L}_{\Delta})$. Lian and Zuckerman
proved that the cohomology classes $H^{\rm abs}_*({\cal L}_{\Delta})$ are
nontrivial if and only if $\Delta \in E$. Recall that $E$ is the set of the
highest weights appearing in the embedding diagram of Verma
modules~(\ref{abdimensions}). It is possible to extend Lian and Zuckerman
consideration and formulate

\begin{thm}\label{AbsTheorem}
If $\Delta\in E$, the dimension of the cohomology space $H^{\rm abs}_*({\cal
L}_{\Delta})$ are given by
$$
\dim H^{\rm abs}_k({\cal L}_{a_n})=\dim H^{\rm abs}_k({\cal
L}_{b_n})=\left\{ \Aligned{
&1, \quad k=-n+1,~-n+3,\ldots,~n-1,\\
&1, \quad k=-n+4,~-n+6,\ldots,~n+2, \\
&0, \quad \text{\rm otherwise} \\
}\right.
$$
where we assume that $n>0$. In the case $n=0$ the dimension of the
cohomology space is $$\dim H^{\rm rel}_k({\cal
L}_{a_0})=\delta_{k,1}+\delta_{k,2}.$$
\end{thm}

\textbf{Proof.} We prove the theorem for $H^{\rm abs}_k({\cal L}_{a_n})$
(the other case can be done in similar manner). Following
\cite{FGZ},\cite{LianZuckerman} one has a long exact sequence containing
relative and absolute cohomology spaces
\begin{equation}
\ldots \stackrel{\gamma_{k-1}}{\longrightarrow} H^{\rm rel}_{k-2}
\stackrel{\alpha_{k-1}}{\longrightarrow} H^{\rm rel}_k
\stackrel{\beta_k}{\longrightarrow} H^{\rm abs}_k
\stackrel{\gamma_{k}}{\longrightarrow} H^{\rm rel}_{k-1}
\stackrel{\alpha_k}{\longrightarrow} H^{\rm rel}_{k+1}
\stackrel{\beta_{k+1}}{\longrightarrow} H^{\rm abs}_{k+1}
\stackrel{\gamma_{k+1}}{\longrightarrow} \label{LongSeq}
\end{equation}
It follows from the exactness of this sequence that $$\dim H^{\rm
abs}_k({\cal L}_{a_n})= \dim \ima(\gamma_k) + \dim \ima(\beta_k)=\dim
\ker(\alpha_k)+\dim H^{\rm rel}_k({\cal L}_{a_n})- \dim
\ima(\alpha_{k-1}).$$ Since the dimension of the cohomology space $\dim
H^{\rm rel}_k({\cal L}_{a_n})$ are known (\ref{LZLiouville}), it is
sufficient to study the maps~$\alpha_k$.

The map $\alpha_* {:} H^{\rm rel}_{*-1} \rightarrow H^{\rm rel}_{*+1}$ is
defined by the action of the ghost operator $c_0$ and then the action of the
BRST charge $Q$. Note that $[Q,c_0]=X$. Therefore the map $\alpha_*$ is
equivalent to the action of the operator $X$. Thus, from Theorem
\ref{newbase} we conclude, that the kernel of the map $\alpha_*$ is spanned
on the cohomology classes of the form $X_+^jO_{a_n}$ ($1\leq j< n$) and
$X^nO_{a_n}$. The image of this map is spanned on the cohomology classes
$X^j O_{a_n}$ ($1\leq j\leq n$). From this consideration the dimension of
cohomology spaces can be easily obtained. $\hfill \square$

The meaning of the long exact sequence (\ref{LongSeq}) is the following. Any
absolute cohomology class can be either a class from $H^{\rm rel}_*({\cal
L}_\Delta)$, or represented by a state of the form $c_0w+w'$, where $w\in
H^{\rm rel}_*({\cal L}_{\Delta})$ and $w'\in C^{\rm rel}_*({\cal
L}_\Delta)$. Some relative cohomology classes are not absolute cohomology
classes. Indeed, one can consider the action of the BRST operator on state
of the form $c_0w+w'$ and obtain
\begin{equation}
Q(c_0w+w')=Xw+Qw'\label{Qc_0}.
\end{equation}
Thus, every relative cohomology class of the form $Xw$ is BRST exact in the
absolute cohomology space. From the Theorem \ref{newbase} it follows, that
the relative cohomology classes
\begin{equation}
O_{a_n},\quad X_+O_{a_n},\quad X^2_+O_{a_n},\quad\ldots\quad X^{n-1}_+O_{a_n}\label{X+O}
\end{equation}
aren't of the form $Xw$ and therefore form a basis in the space $H^{\rm
rel}_*({\cal L}_{a_n})\cap H^{\rm abs}_*({\cal L}_{a_n})$. In order to
extend the set~(\ref{X+O}) to basis we need to add some states in form
$c_0w+w'$, where $w\in H^{\rm rel}_*({\cal L}_{a_n})$ and $w'\in C^{\rm
rel}_*({\cal L}_{a_n})$. It follows from (\ref{Qc_0}) that if $Q(c_0w+w')=0$
then $Xw=0$ in $H^{\rm rel}_*({\cal L}_{a_n})$. By Theorem~\ref{newbase}
such $w$ is a linear of combination of cohomology classes
$X_+O_{a_n},~X^2_+O_{a_n},~\ldots~,X_+^{n-1}O_{a_n}, X^nO_{a_n}$. Then all
additional basic states have form
$c_0X_+O_{a_n},~c_0X^2_+O_{a_n},~\ldots~,c_0X_+^{n-1}O_{a_n},~c_0X^nO_{a_n}$
modulo $C^{\rm rel}_*({\cal L}_{a_n})$.

Let us introduce a new operator
\begin{equation}
Y=\frac1{12}\sum\limits_{i+j+k=0}(i-j)(j-k)(k-i)c_ic_jc_k\nn
\end{equation}
with the ghost number 3. It is easy to check, that this operator commutes
with the BRST charge $Q$ and thus acts on the cohomology space. Moreover, if
$w\in H^{\rm rel}_*({\cal L}_{a_n})$, then $Yw=c_0X_+w+w'$ with $w'\in
C^{\rm rel}_*({\cal L}_{a_n})$.
%(Note the difference between (\ref{Y}) and (\ref{tildeY})).

It follows from previous consideration that the set (\ref{X+O}) can be
expanded to form a basis in the absolute cohomology space $H^{\rm abs}_*
({\cal
 L}_{a_n})$ by adding the following cohomology classes
$$
YO_{a_n},~YX_+O_{a_n},~YX^2_+O_{a_n},~\ldots~ YX_+^{n-1}O_{a_n}.
$$

Indeed, $YO_{a_n}=c_0X_+O_{a_n}$,
$YX_+O_{a_n}=c_0X^2_+O_{a_n}$,...,$YX^{n-2}_+O_{a_n}=c_0X^{n-1}_+O_{a_n}$
modulo $C^{\rm rel}_*({\cal L}_{a_n})$. It remains to prove that
$YX_+^{n-1}O_{a_n} \neq 0$ in cohomology space $H^{\rm abs}_*({\cal
L}_{a_n})$. This can be proven by duality arguments similar to remark
\ref{duality}

Our discussion of the cohomology classes of the absolute BRST complex
$C^{\rm abs}_*({\cal L}_{a_n})$ can be summarized by the following diagram

\begin{center}
\begin{picture}(350,150)
\put(40,90){$O_{a_n}$} \put(55,94){\vector(1,0){35}}
\put(55,94){\vector(4,-3){65}} \put(65,97){$X_+$} \put(80,60){$Y$}

\put(95,90){$X_+O_{a_n}$} \put(125,40){$YO_{a_n}$}

\multiput(120,94)(65,0){2}{\vector(1,0){30}}
\multiput(150,43)(75,0){2}{\vector(1,0){30}}

\multiput(128,98)(65,0){2}{$X_+$} \multiput(160,47)(75,0){2}{$X_+$}

\put(155,90){$X^2_+O_{a_n}$} \put(185,40){$YX_{+}O_{a_n}$}

\put(260,40){$\ldots$} \put(220,90){$\ldots$}

\put(235,94){\vector(1,0){30}} \put(275,43){\vector(1,0){30}}

\put(243,98){$X_+$} \put(285,47){$X_+$}

\put(310,40){$YX^{n-1}_+O_{a_n}$} \put(270,90){$X^{n-1}_+O_{a_n}$}

\put(0,0){$N^g=$} \put(115,0){$-n+4,$} \put(185,0){$-n+6,$}
\put(260,0){$\ldots$} \put(320,0){$n+2.$}

\put(0,130){$N^g=$} \put(35,130){$-n+1,$} \put(90,130){$-n+3,$}
\put(150,130){$-n+5,$} \put(220,130){$\ldots$} \put(280,130){$n-1.$}
\end{picture}
\end{center}
%\begin{figure}[htbp]\center
%\includegraphics[width=90mm]{operatorsRel.eps}\label{basis2}
%\end{figure}
%\begin{figure}[htbp]\center
%\includegraphics[width=85mm]{basisAbs.eps}
%\caption{The basis in the space $H_*^{abs}{\cal L}_{a_n}$}\label{basis}
%\end{figure}
Let $\gamma_n$ be either $a_n$ or $b_n$. It is convenient to denote the
basic cohomology classes in the following form
\begin{eqnarray}
O_{\gamma_n}^{i}=(n-i-1)!X_+^i O_{\gamma_n} \qquad
N_{\gamma_n}^{i}=(n-i-1)!YX_+^i O_{\gamma_n}, \quad \text{where} \quad 0\leq
i<n.\label{ON}
\end{eqnarray}
We will see that this notation simplify the structure constants of the
operator algebra. We clearly have
\begin{eqnarray}
X_+O_{\gamma_n}^{i}=(n-i-1)O_{\gamma_n}^{i+1},\qquad
X_+N_{\gamma_n}^{i}=(n-i-1)O_{\gamma_n}^{i+1}.\label{ONX+}
\end{eqnarray}

Ghost numbers for these cohomology classes are given by
$$
N^g(O_{\gamma_n}^{i})=2i-n+1,\qquad N^g(N_{\gamma_n}^{i})=2i-n+4.
$$

%The ghost numbers in Fig.(\ref{basis}) are symmetric with respect to $3/2$. Thus, in the space $H^{abs}{\cal L}_{a_n}$ there exists well defined genus zero two point function
%\begin{equation}
%\la O^{(i,0)}_{a_{n+1}}O^{(j,1)}_{a_{n+1}}\ra=\delta_{j+i,n}\label{2poinFunction}
%\end{equation}
%where we assume that the state are normalized in appropriate way.

%We emphasize, that the genus zero two point function exist only in the whole cohomology space %$H^{abs}_{*}{\cal L}$, not in $H^{rel}_{*}{\cal L}$ or $H^{abs}_{*}{\cal L}\cap H^{rel}_{*}{\cal L}$.

\subsection{Operator Algebra}

The operator product expansion provides the ring structure (\ref{Ring}) on
the absolute cohomology space. By the basic assumptions this ring is
associative and commutative. There is an unit element in the ring namely the
identity operator $O_{a_1}^{0}(z)={\mathbb I}(z)$.

As we will show an operator algebra in the absolute cohomology space is
almost determined by the operators $X_+$ and $Y$. Thus it is useful to
consider this operators first. Due to operator-state correspondence any
operator that acts on the Hilbert space of states has an image acting in the
space of local operators. One can show that the image of the operator $X_+$
is given by the contour integral

\eq{ X_+=-\oint dz c(z)\p^3c(z),\label{X+Phys} } where we omit BRST exact
terms. As a consequence of this representation we conclude that the operator
$X_+$ differentiates the product of any two local operators, that is \eq{
X_+(O_1\cdot O_{2})=(X_+O_1)\cdot O_2+O_1\cdot(X_+O_2).\label{X+diff} }

Now consider the operator $Y$. It is straightforward to check that on the
space of local operators the application of the $Y$ is equivalent to the
zero mode of the operator product with $\frac12{:}c\p c\p^2c{:}$, i.e
$$YO(0)=\frac12 Res_{z=0}\left(\frac{{:}c(z)\p c(x)\p^2c(z){:}O(0)}{z}\right) $$
modulo BRST exact terms. Note that $\frac12{:}c\p c\p^2c{:}$ is local
operator corresponding to the cohomology class~$N_{a_1}^{0}$:
$$
N_{a_1}^{0}=\frac12{:}c\p c\p^2c{:}.
$$
Therefore on the cohomology space we have \eq{ YO=N^{0}_{a_1}\cdot
O.\label{YPhys}}

The study of the operator algebra is simplified due to fusion rules. First
we consider the case when operators are of the form $O_{a_n}^i$. The
corresponding states form a basis in $H^{\rm rel}_*({\cal L}_{a_n})\cap
H^{\rm abs}_*({\cal L}_{a_n})$. Taking into account the fusion rules for
degenerate Virasoro representations \cite{BPZ} and the ghost number
conservation, we have \eq{ O_{a_{k+1}}^{i}\cdot
O_{a_{l+1}}^{j}=\sum_{n=0}^{i+j}\lambda_n
O_{a_{k+l+1-2n}}^{i+j-n},\label{OPEabsGen} } with some structure constants
$\lambda_n$ depending on all indexes $i,j,k,l$.

The operators $X_+$ and $Y$ almost determine structure constants of the
operator algebra due to the next two propositions.
\begin{prop}\label{Proppolyn}
Suppose that
\begin{equation}
O_{a_{k+1}}^{0}\cdot O_{a_{l+1}}^{0}=O_{a_{k+l+1}}^{0}.\label{hixhi}
\end{equation}
Then the subring $H^{\rm rel}_*({\cal L}_{a_n})\cap H^{\rm abs}_*({\cal
L}_{a_n})$ is isomorphic to the polynomial ring in two generators
$O_{a_2}^0$ and $O_{a_2}^1$. Moreover, we have \eq{
O_{a_{n+1}}^i=(O_{a_2}^0)^{n-i}(O_{a_2}^1)^i \label{OGener}}
\end{prop}
In the assumption of this proposition we suppose that the operator product
is non-degenerate, while the corresponding coefficient (see \ref{OPEabsGen})
can be removed by certain normalization of the lowest cohomology
classes~$O_{a_n}$.

\textbf{Proof of Proposition \ref{Proppolyn} }

It is enough to prove~(\ref{OGener}). For $n=0$ the equation~(\ref{OGener})
is equivalent to the fact that $O_{a_1}^{0}$ is a unit element. For $n=1$
the equation~(\ref{OGener}) is obvious. From the assumption~(\ref{hixhi})
follows that \eq{ O_{a_{n+1}}^0=O_{a_n}^0 \cdot O_{a_2}^0=O_{a_{n-1}}^0
\cdot O_{a_2}^0 \cdot O_{a_2}^0 = \ldots= (O_{a_2}^0)^n \label{O^0}}

Applying $X_+$ to both sides and using~(\ref{X+diff}),~(\ref{ONX+}) we get
$$nO_{a_{n+1}}^1=n(O_{a_2}^0)^{n-1}\cdot O_{a_2}^1, \qquad  O_{a_{n+1}}^1=(O_{a_2}^0)^{n-1}\cdot
O_{a_2}^1.$$ Applying $X_+$ to both sides again and
using~(\ref{X+diff}),~(\ref{ONX+}) we get
$$(n-1)O_{a_{n+1}}^2=(n-1)(O_{a_2}^0)^{n-2}\cdot O_{a_2}^1 \cdot O_{a_2}^1
\qquad O_{a_{n+1}}^2=(O_{a_2}^0 )^{n-2}\cdot (O_{a_2}^1)^2.$$ Similarly,
applying $X_+$ $i$ times to both sides of~(\ref{O^0}) we get~(\ref{OGener}).
 $\hfill \square$

It is interesting to compare our results with those of Kanno and Sarmadi
\cite{KS}, where irreducible modules in the Liouville sector are replaced by
free field (or Feigin-Fuchs) modules. As was mentioned in
remark~\ref{correspon} the cohomology classes from \cite{KS} correspond to
our relative cohomology classes with the highest and the lowest ghost
numbers. For example $w^n$ in their notation corresponds to $O^0_{a_{n+1}}$
with $n \geq 0$. The results of \cite{KS} confirm that the products of the
cohomology classes of the form $O^{0}_{a_{n+1}}$ are non-degenerate and the
assumption of Proposition \ref{Proppolyn} is satisfied.

The structure constants of the operator algebra on the space
$\oplus_{n>0}H^{\rm abs}_*({\cal L}_{a_n})$ are given in
Proposition~\ref{PropStructConst}.

\begin{prop}\label{PropStructConst}
Under the assumption of Proposition~\ref{Proppolyn} we have
\begin{equation}
\begin{gathered}
O_{a_{k+1}}^{k-i}\cdot O_{a_{l+1}}^{l-j}=O_{a_{k+l+1}}^{k+l-i-j},\qquad N_{a_{k+1}}^{k-i}\cdot O_{a_{l+1}}^{l-j}=N_{a_{k+l+1}}^{k+l-i-j}, \\
N_{a_{k+1}}^{k-i}\cdot N_{a_{l+1}}^{l-j}= 0.\label{StrConst}
\end{gathered}
\end{equation}
\end{prop}

\textbf{Proof of Proposition \ref{PropStructConst} }

The first equality evidently follows from~(\ref{OGener}). Let us multiply
the first equality by $N_{a_1}^0$. By~(\ref{YPhys}) and~(\ref{ON}) we obtain
$$N_{a_{k+1}}^{k-i}\cdot O_{a_{l+1}}^{l-j}=N^{0}_{a_1} \cdot
O_{a_{k+1}}^{k-i}\cdot O_{a_{l+1}}^{l-j}=N^{0}_{a_1} \cdot
O_{a_{k+l+1}}^{k+l-i-j}= N_{a_{k+l+1}}^{k+l-i-j}.$$ The last statement in
(\ref{StrConst}) easily follows from the equalities
$N_{a_{k+1}}^{k-i}=YO_{a_{k+1}}^{k-i}=N_{a_1}^{0} \cdot O_{a_{k+1}}^{k-i}$
and $N_{a_1}^{0} \cdot N_{a_1}^{0}=YN_{a_1}^{0}(z)=0$. $\hfill \square$

Note, that we consider an operator algebra for the states from the subspace
$\oplus_{n>0}H^{\rm abs}_*({\cal L}_{a_n})$ only. It is sufficient, because
there is an isomorphism between the cohomology spaces $H^{\rm abs}_*({\cal
L}_{a_n})$ and $H^{\rm abs}_*({\cal L}_{b_n})$. This isomorphism is realized
by the operator product with
$$
O_{b_1}^{0}(z)=(\p+\frac{2}{3}:bc:){\Phi}_{b_1}(z),
$$
where ${\Phi}_{b_1}(z)={\Phi}_{1,2}(z)$ is a Liouville primary field
corresponding to the state $|{\cal L}_{b_1}\ra$ from the Hilbert space.
Indeed, by the fusion rules for degenerate Virasoro representations
\cite{BPZ} and the ghost number conservation, we have \eq{ O_{b_1}^{0}\cdot
O_{a_{n+1}}^{i}=\lambda_{b_1,a_{n+1}}^{i}O_{b_{n+1}}^{i},\quad
O_{b_1}^{i}\cdot
O_{b_{n+1}}^{i}=\lambda_{b_1,b_{n+1}}^{i}O_{a_{n+1}}^{i},\label{Isom} }
where $\lambda_{b_1,a_{n+1}}^{i}$ and $\lambda_{b_1,b_{n+1}}^{i}$ are the
structure constants. To show that these constants are not equal to~0, we
multiply both sides of the first equation in~(\ref{Isom}) by the operator
$O_{b_1}^{0}$. Taking into account the associativity of the operator
algebra, we have
$$(O_{b_1}^{0})^2\cdot O_{a_{n+1}}^{i} =
\lambda_{b_1,a_{n+1}}^{i}O_{b_1}^{0}\cdot
O_{b_{n+1}}^{i}=\lambda_{b_1,a_{n+1}}^{i}\lambda_{b_1,b_{n+1}}^{i}O_{a_{n+1}}^{i}
$$
Since \eq{(O_{b_1}^{0})^2=-14/9C_{(1,2),(1,2)}^{(1,1)}{\mathbb
I}(z),\label{Ob10}} where $C_{(1,2),(1,2)}^{(1,1)}$ is the Liouville
structure constant (\cite{DO},\cite{ZamZam}), we conclude that
$$\lambda_{b_1,a_{n+1}}^{i}\lambda_{b_1,b_{n+1}}^{i}=-14/9C_{(1,2),(1,2)}^{(1,1)}\neq
0.$$

It is possible to renormalise local operators $O_{b_n}^0$ such that
$$
O_{b_1}^{0}\cdot O_{a_{n+1}}^{0}=O_{b_{n+1}}^{0}.
$$
Applying operators $X_+$ and $Y$ to both sides we get \eq{ O_{b_1}^{0}\cdot
O_{a_{n+1}}^{i}=O_{b_{n+1}}^{i}, \qquad O_{b_1}^{0}\cdot
N_{a_{n+1}}^{i}=N_{b_{n+1}}^{i} \label{Hab}}

Using Proposition~\ref{PropStructConst} and formulae (\ref{Ob10}),
(\ref{Hab}) one can easily calculate operator product of any two local
fields.

\section{Discussion}

One of the main aims of this paper is to clarify the difference between
absolute and relative cohomology classes. For mathematical reasons, it is
convenient to compute relative cohomology first and then pass to absolute
cohomology. In some cases (for example in the case of free field modules in
the Liouville sector \cite{KS}) there is am isomorphism $H^{\rm abs}_k \cong
H^{\rm rel}_k+c_0 H^{\rm rel}_{k-1}$. In particular, any relative cohomology
class is also an absolute cohomology class. In our case the relation between
absolute and relative cohomology classes is more complicated.

It was proved in Proposition~\ref{Proppolyn} that the structure of the
operator algebra in the space $H^{\rm rel}_*({\cal L}_{a_n})\cap H^{\rm
abs}_*({\cal L}_{a_n})$ is isomorphic to the algebra of polynomials in two
variables ${\mathbb C}[a,b]$ and the isomorphism is realized by
\begin{equation}
O_{a_{n+1}}^i\mapsto a^{n-i}b^i.\nn
\end{equation}
It is well known that $sl_2$ acts on ${\mathbb C}[a,b]$. Thus one can expect
that $sl_2$ acts in the space $H^{\rm rel}_*({\cal L}_{a_n})\cap H^{\rm
abs}_*({\cal L}_{a_n})$. It is easy to check, that the operator $X_{+}$
corresponds to the $sl_2$ increasing generator
\begin{equation}
X_+\mapsto b\frac{\p}{\p a}.\nn
\end{equation}
We may expect that there exists an operator $X_-$ corresponding to the
$sl_2$ decreasing generator
\begin{equation}
X_-\mapsto a\frac{\p}{\p b}.\nn
\end{equation}
A construction of this operator is an open problem. It seems that there is
no decreasing operator $X_-$ such that $[Q,X_-]=0$ and $X_-$ acts nonzero on
the cohomology space. It is expected that there exists operator $X_-$ such
that $[Q,X_-] \neq 0$, but $X_-$ acts on certain representatives of $H^{\rm
rel}_*({\cal L}_{a_n})\cap H^{\rm abs}_*({\cal L}_{a_n})$. This is similar
to $sl_2$ action on the space of harmonic forms on a Kahler manifold
\cite{GH}.

In ref. \cite{LZ-BV} Lian and Zuckerman have recognized that absolute
cohomology has a structure of Gerstenhaber algebra. In other words they
define a bracket $\{u,v\}$. This bracket differentiates an operator product
and provides the structure of a Lie algebra on the absolute cohomology space
. It would be interesting to calculate this bracket in our case. The
simplest example is $\{N_{a_1}^{0},O \}=X_+O$. This equality is equivalent
to (\ref{X+Phys}) and explains the fact that $X_+$ differentiates operator
product.

\section{Acknowledgments}

We are grateful to A.~Belavin for posing the problem and constant attention
to our work. We are also grateful to O.~Bershtein, B.~Feigin, M.~Lashkevich,
A.~Losev, D.~Polyakov and Y.~Pugai for useful discussions. The work was
supported in part by the Russian Foundation of Basic Research under the
grant RFBR 07-02-00799 and initiative interdisciplinary project grant
09-02-12446-ofi\_m and by the Russian Ministry of Science and Technology
under the Scientific Schools grant 3472.2008.2. This research was held
within the bounds of Federal Program "Scientific and Scientific-Pedagogical
personnel of innovational Russia" on 2009-2013 y., goskontrakt N P1339.

\section{Appendix}

In this Appendix we give an explicit form of some states in the relative
cohomology space. Explicit form of several the highest cohomology classes
are given in (\ref{HighestCo}). Here we consider the other states. These
examples are obtained by recursion procedure described above.
 Let us start from the

\textbf{Embedding level 1.}\nopagebreak

Let us consider the cohomology space $H^{\rm rel}_{*}({\cal L}_{a_1})$. It
is easy to check that the cohomology class of the ghost number 0 has the
following form
$$
O_{a_1}^{a_0}=H_{a_1}^{a_0}\Psi^{\cal L}_{a_1}=b_{-1}\Psi^{\cal L}_{a_1}
$$
One can show that $Q(O_{a_1}^{a_0})=L_{-1}\Psi^{\cal L}_{a_1}=0$, since
$L_{-1}|V_{a_1}\ra$ is a singular vector in the Verma module $V_{a_1}$ and,
therefore, in the irreducible module $L_{-1}|{\cal L}_{a_1}\ra=0$.

In the space $H^{\rm rel}_{*}({\cal L}_{b_1})$ the cohomology class of the
ghost number 0 is
$$
O_{b_1}^{a_0}=H_{b_1}^{a_0}\Psi^{\cal
L}_{b_1}=\left(b_{-1}L_{-1}+\frac{2}{3}b_{-2}\right)\Psi^{\cal L}_{b_1}
$$
It is easy to check that $Q(O_{b_1}^{a_0})=(L^2_{-1}+(2/3)L_{-2})\Psi^{\cal
L}_{b_1}=0$, since $(L^2_{-1}+(2/3)L_{-2})|{V}_{b_1}\ra=0$ is a singular
vector in the Verma module $V_{b_1}$ and, therefore, in the irreducible
module $(L^2_{-1}+(2/3)L_{-2})|{\cal L}_{b_1}\ra=0$.

\textbf{Embedding level 2.}\nopagebreak

Here we only consider the cohomology space $H^{\rm rel}_{*}({\cal
L}_{a_2})$. The highest cohomology class is given in~(\ref{HighestCo}). Let
us consider the rest of them.

It is straightforward to check that cohomology classes of the ghost number 1
can be written as follows
\begin{equation}
\begin{aligned}
O_{a_2}^{a_1}=H_{a_2}^{a_1}\Psi^{\cal L}_{a_2}&=\Bigl(-c_{-1}b_{-1}L_{-1}^3-
\frac{20}{3}c_{-1}b_{-2}L_{-1}^2-4c_{-1}b_{-2}L_{-2}-\frac{52}{3}c_{-1}b_{-3}L_{-1}+
3c_{-2}b_{-1}L_{-1}^2+\\
&+\frac{64}{3}c_{-3}b_{-1}L_{-1}-\frac{76}{3}c_{-1}b_{-4}+20c_{-3}b_{-2}+
\frac{44}{3}c_{-4}b_{-1}\Bigr)\Psi^{\cal L}_{a_2}\\
O_{a_2}^{b_1}=H_{a_2}^{b_1}\Psi^{\cal L}_{a_2}&=\left(-c_{-2}b_{-1}L_{-1}^2
-6c_{-2}b_{-2}L_{-1}+4c_{-3}b_{-1}L_{-1}-12c_{-2}b_{-3}+16c_{-4}b_{-1}\right)\Psi^{\cal
L}_{a_2}.\nn
\end{aligned}
\end{equation}

Indeed, one can check that $Q(O_{a_2}^{a_1})=D_{b_1,a_2}\Psi^{\cal
L}_{a_2}=0$ and $Q(O_{a_2}^{b_1})=D_{b_1,a_2}\Psi^{\cal L}_{a_2}=0$. The
explicit forms of the operators $D_{a_1,a_2}$ and $D_{b_1,a_2}$ can be found
in \cite{FF2}.

The cohomology class with the ghost number -1 is
\eq{O_{a_2}^{a_0}=\left(-\frac{2}{3}b_{-2}b_{-1}(L_{-1}^2+6L_{-2})+\frac{2}{3}b_{-3}b_{-1}L_{-1}-
\frac{4}{3}b_{-4}b_{-1}+4b_{-3}b_{-2}\right)\Psi^{\cal L}_{a_2},
\label{Oa_2a_0} } and one can check that
$$ Q(O_{a_2}^{a_0})=H_{a_{1}}^{a_0}
D_{a_1,a_2}\Psi^{\cal L}_{a_2} - H_{b_1}^{a_0}D_{b_1,a_2}\Psi^{\cal
L}_{a_2}=0$$
 which is in agreement with the result of our recursive
construction procedure.

Due to remark~\ref{correspon} the formula~(\ref{Oa_2a_0}) for the lowest
cohomology~(\ref{Oa_2a_0}) have an analogue is the free field approach.
Indeed this formula is equivalent to the formula~(2.9)~in~\cite{Govin2}
obtained by the different method.

\end{document}